# Viskositas: Viscosity Prediction of Multicomponent Chemical Systems

Patrick dos Anjos[1]
*Federal Institute of Espírito Santo (IFES), Vitória, ES, Brazil*

## Abstract

Viscosity in the metallurgical and glass industry plays a fundamental role in its production processes, also in the area of geophysics. As its experimental measurement is financially expensive, also in terms of time, several mathematical models were built to provide viscosity results as a function of several variables, such as chemical composition and temperature, in linear and nonlinear models. A database was built in order to produce a nonlinear model by artificial neural networks by variation of hyperparameters to provide reliable predictions of viscosity in relation to chemical systems and temperatures. The model produced named *Viskositas* demonstrated better statistical evaluations of mean absolute error, standard deviation and coefficient of determination ($R^2$) in relation to the test database when compared to different models from literature and 1 commercial model, offering predictions with lower errors, less variability and less generation of outliers.

## 1 Introduction

Viscosity is studied in phenomena that involve interaction between particles, such as in the desulfurization of steels [1] and in the treatment of inclusions in the steels in the tundish [2], as well as in the consumption of flux during the continuous casting of steels [3] and in the calculation of the thermal conductivity of aluminosilicates [4]. In glass rheology, viscosity is required for different manufacturing actions, having restricted values for practices such as refining and homogenization [5] and in the knowledge of multicomponent silicate mixtures of rock forming compositions [6].

Viscosity can be measured experimentally through several methods, from falling sphere, rotating cylinder, rod elongation viscosimeters to squeeze film rhemoter. But experimental methods have inherent errors about the procedural process itself, incurring errors that can lead to improper measurement, distorting the results. There may be 5-20% accuracy errors and 1-5% precision errors [7]. The experimental process of measuring viscosity is expensive both in terms of economics, since the equipment is expensive, in terms of time, since measuring viscosity requires homogenization of the temperature of the sample to be tested, and in terms of operator experience, since the minutiae of the experimental process can distort the final results [8-10].

---
1  E-mail: patrick.dosanjos@outlook.com

Numerical methods from experimental data and/or semi-empirical considerations have been developed to provide satisfactory results under different conditions at different chemical compositions and temperatures [7]. But most models have restricted application, with validation in chemical composition, temperature and viscosity values that limit their application. One of the ways to overcome the limitation of the applicability of these models is to expand the database during mathematical modeling and apply computational models, such as artificial neural networks.

Artificial neural networks (ANN) are computational models capable of establishing mathematical relationships in order to approach an optimal value, by updating these mathematical relationships in a finite number of interactions. Artificial neural networks are widely used in classification, resulting in discrete variables for image processing and recognition [11] with image-based search engines and biomedical diagnosis [12] and regression where the result is a continuous variable for the prediction of mechanical properties in microalloyed steels [13] and nickel superalloys [14] and in the prediction of viscosity in metallurgical slags [15] and oxide liquids [16].

This work aims to implement a neural network capable of predicting viscosity through the chemical composition of multicomponent chemical systems and temperature through a literature database to establish a nonlinear model.

## 2 Materials and Methods

### 2.1 Preprocessing Database

The database was built from 2 international scientific articles and 1 bachelor thesis. The references, the amount of data, chemical system, temperature (°C or K) and viscosity (Pa·s), can be seen in Table 1.

Table 1 — Amount of data and the chemical system used in database.

| Reference | Data | Chemical system |
|---|---|---|
| Duchesne et al. [17] | 4124 | $SiO_2$-$Al_2O_3$-$Fe_2O_3$-FeO-Fe-CaO-MgO-$Na_2O$-$K_2O$-$Li_2O$-MnO-$TiO_2$-$B_2O_3$-$P_2O_5$-NiO-$ZrO_2$-$CaF_2$-$SO_3$-$Cr_2O_3$-$V_2O_5$ |
| Chen et al. [18] | 1892 | CaO-$SiO_2$-MgO-$Al_2O_3$-$Fe_nO$*-$R_2O$**-$TiO_2$-$Mn_nO$* |
| Anjos [19] | 1019 | CaO-$SiO_2$-MgO-$Al_2O_3$-$TiO_2$-MnO-FeO-$CaF_2$-$Na_2O$-$Li_2O$-$B_2O_3$-$K_2O$-$ZrO_2$-$Fe_2O_3$. |

* $Fe_nO$ e $Mn_nO$ equals FeO and MnO
** $R_2O$ = $Na_2O$ + $K_2O$ with %$Na_2O$ = %$K_2O$

The chemical system is $SiO_2$-$Al_2O_3$-$Fe_2O_3$-FeO-Fe-CaO-MgO-$Na_2O$-$K_2O$-$Li_2O$-MnO-$TiO_2$-$B_2O_3$-$P_2O_5$-NiO-$ZrO_2$-$CaF_2$-$SO_3$-$Cr_2O_3$-$V_2O_5$. The distribution of viscosity data ($\eta$), in logarithm base-10, can be seen in Figure 1 (a). It is possible to see the mean, arithmetic mean, ($\mu$) and standard deviation ($\sigma$) of the viscosity distribution ($\log_{10} \eta$), 0.3869 and 1.5661, respectively. The preprocessing was carried out to establish the degree of depolymerization (NBO/T), the liquidus temperature (Tliq) and the standardization by the z-score (z).

The depolymerization degree parameter (NBO/T) is used to establish the depolymerization in different chemical systems. Electrical resistivity, diffusion coefficient, thermal conductivity and viscosity are highly dependent on the degree of polymerization $Q$ ($Q = 4 - NBO/T$) [20]. The NBO/T can be calculated using Equation 1.

$$NBO/T = \frac{2\left(\sum X_{MO} + \sum X_{M_2O} - \sum X_{M_2O_3}\right)}{\sum X_{MO_2} + 2\sum X_{M_2O_3}} \quad (1)$$

$X_{MO}$ = XCaO + MgO + …, $X_{M2O}$ = $XNa_2O$ + $XLi_2O$ + …, $X_{M2O3}$ = $XAl_2O_3$ + $XB_2O_3$ + … e $X_{MO2}$ = $XSiO_2$ + $XTiO_2$ + … e $X_{MO}$ equivalent to the molar fraction (X) of the chemical species MO. With the parameter NBO/T, the pre-processing of the database was carried out for the chemical composition data, in relation to the NBO/T, are not considered outliers. Therefore, the NBO/T limits were established as described in Equation 2.

$$F_{quartile} - 1.5(T_{quartile} - F_{quartile}) < NBO/T < T_{quartile} + (T_{quartile} - F_{quartile}) \qquad (2)$$

$F_{quartile}$ is equivalent to the first quartile and $T_{quartile}$ to the third quartile of the NBO/T distribution. The liquidus temperature (Tliq) is the temperature at which a material becomes fully liquid [20]. This preprocessing is necessary to have in the mathematical and computational modeling only fully liquid fluids, considered Newtonian, which can be governed by the Arrhenius [21-22], Weymann-Frenkel [23] or Vogel-Fulcher-Tammann (VFT) equation [24]. Since the database does not provide shear rate and/or stress data, it would not be inherent to non-Newtonian fluid modeling [25-26]. In Equation 3, the liquidus temperature can be seen as a function of the chemical composition of different materials [20].

$$T_{liq}(K) = 1473 - 1.518\%SiO_2 + 2.59\%CaO + 1.56\%Al_2O_3 - 17.1\%MgO - 9.06\%Na_2O - 6.0\%K_2O + 18\%Li_2O + 4.8\%F - 9.87\%FeO - 2.12\%MnO \qquad (3)$$

%$A$ is equivalent to mass percentage of $A$ chemical specie. The database was preprocessed in order to keep only chemical compounds with the experimental viscosity measurement temperature above their liquidus temperature. Similar to NBO/T, database viscosity preprocessing was performed to remove outliers (Equation 2). The viscosity in $\log_{10} \eta$ in the database varies between –3 and 8. The viscosity distribution ($\log_{10} \eta$) can be seen in Figure 1 (b).

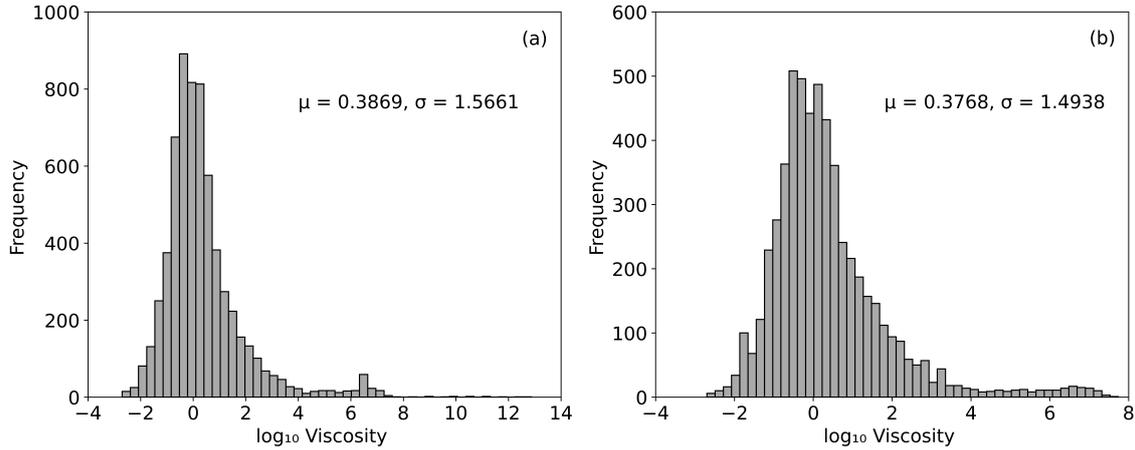

Figure 1 — Viscosity distribution ($\log_{10} \eta$) in the database before preprocessing (a) and after preprocessing (b).

In Figure 1 (b) it can be seen that the mean ($\mu$) and standard deviation ($\sigma$) of viscosity ($\log_{10} \eta$) of the preprocessing database have values of 0.3768 and 1.4938 respectively. After preprocessing the viscosity was transformed into logarithm ($\log_{10}$) in order to reduce the dimensionality and adapting to the mathematical modeling of viscosity [21-24]. Any duplicates in the database were removed. The limits of the database submitted to the artificial neural network can be seen in Table 2.

The database was partitioned into a database of predictive variables chemical composition (%mass) and temperature (K), and into a database of target variable $\log_{10} \eta$ ($\eta$ in Pa·s). After the database was standardized by the z-score equation (Equation 4). Standardization via z-score modifies the distribution of the labels in the database to obtain a mean and a standard deviation equal to 0 and 1 of these labels.

$$z_i = \frac{x_i - \mu_i}{\sigma_i} \qquad (4)$$

where $x_i$, $z_i$, $\mu_i$ and $\sigma_i$ are equivalent to a data from the data set x, its z-score, the mean and the standard deviation of this data set respectively. This preprocessing was performed in relation to the predictor variables. The z-score is made to reduce the dimensionality of the variables, standardizing their distribution, helping in the artificial neural network development process by reducing a parameter of the neural network, the bias [16]. Figure 2 shows the temperature distribution (K) of the database after preprocessing. As mentioned earlier, it can be seen in Figure 2 that the temperature in the database after the application of the z-score has a mean and standard deviation equal to 0 and 1, respectively. After all the pre-processing, the database that initially had a total of 7035 different chemical composition, temperature and viscosity data after preprocessing had a total of 5659 different data.

In the development of an artificial neural network, training and test databases are usually used, and a validation database can also be used [11-17]. The training database is used to obtain the best parameter ratio of the artificial neural network, the validation database assists in the artificial neural network evaluation process during training, and the test database aims at the evaluation of the artificial neural network on unknown data. To this end, the database was partitioned into: 81% for the training database; 9% for the validation database; 10% for the test database. The validation database was *not* used to train the neural network and the test database was *not* used during the training of the neural network.

Table 2 — Maximums and minimums of preprocessed database labels ($3.00 \leq \eta \leq 8.00$ ($\log_{10} \eta$), $\eta$ in Pa·s).

|      | CaO*  | $SiO_2$ | MgO   | $Al_2O_3$ | $TiO_2$ | MnO   | FeO   | $CaF_2$ | $Na_2O$ | $Li_2O$ |
|------|-------|---------|-------|-----------|---------|-------|-------|---------|---------|---------|
| Min. | 0.00  | **      | **    | **        | **      | **    | **    | **      | **      | **      |
| Max. | 78.00 | 100.00  | 55.58 | 100.00    | 49.99   | 72.25 | 83.49 | 34.60   | 35.79   | 20.00   |
|      | $B_2O_3$ | $K_2O$ | $ZrO_2$ | $Fe_2O_3$ | $P_2O_5$ | NiO | $SO_3$ | $Cr_2O_3$ | $V_2O_5$ | T (K)   |
| Min. | 0.00  | **      | **    | **        | **      | **    | **    | **      | **      | 1152.15 |
| Max. | 31.02 | 48.00   | 1.00  | 85.10     | 4.11    | 1.17  | 2.02  | 3.43    | 7.18    | 2755.15 |

* chemical composition in mass percentage (%mass).
** 0.00

## 2.2 Artificial Neural Network

The algorithm chosen to perform the viscosity prediction using predictive variables chemical composition and temperature was the artificial neural network. Many works were carried out to demonstrate the aspects of artificial neural networks, their construction, applicability and different architectures [27-31]. The main aspects that led to this choice were the availability of high-performance and quality libraries for the development of artificial neural networks, such as scikit-learn [32], PyTorch [33] and TensorFlow [34], a vast community about the use of optimization and development of artificial neural networks and the universal approximation theorem [35-38].

Cybenko [35] demonstrated that the approximation of continuous functions can be performed through artificial neural networks with the sigmoid activation function. Hornik et al. [36] established that artificial neural networks with arbitrary activation functions are able to approximate mathematical functions. Leshno et al. [37] demonstrated that neural networks can approximate continuous functions if and only if the activation functions applicable in the neural network are non-polynomial. Lu et al. [38] demonstrated that there is a fully-connected ReLU neural network F that approximates a wide class of functions $f:R^n \rightarrow R$ with error $\varepsilon > 0$.

Hyperparameters of artificial neural networks are variables introduced during training (e.g. number of layers, number of neurons, activation functions [16]) and parameters are the unknowns generated by the neural network after training (e.g. weights, bias). One of the most complex hyperparameters of artificial neural networks is the minimum width of the hidden layer (hidden layer can be composed of one layer or several hidden layers) of the artificial neural network to obtain the universal approximation theorem. Several works were carried out to obtain the minimum width to reach the universal approximation [38-40]. Kidger and Lyons [39] developed an equation establishing the upper limit on the minimum amount of neurons in the hidden layer in artificial neural networks with arbitrary depth (Equation 5).

$$w_{min} \leq d_x + d_y + 1 \tag{5}$$

where $w_{min}$ is the minimum width, $d_x$ is the input set cardinality and $d_y$ the output set cardinality. As the amount of neural network input data is 20 (chemical composition and temperature) and the amount of output data is 1 ($\log_{10} \eta$), the minimum width with arbitrary depth is less than or equal to 22. The hyperparameters introduced during neural network training can be seen in Table 3.

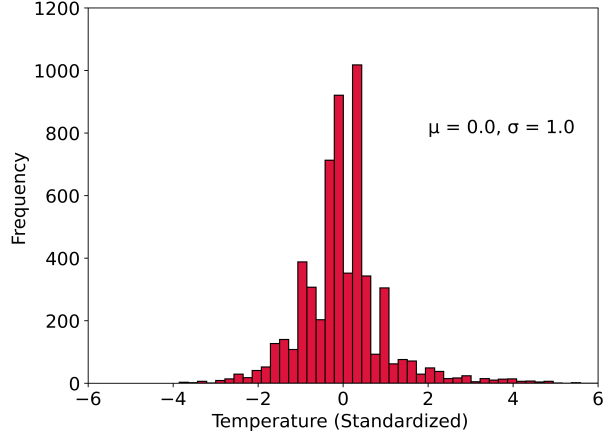

Figure 2 — Temperature distribution (K) in the database after preprocessing.

Experiments were carried out to obtain the minimum loss and minimum metrics. The EarlyStopping technique was used in order to avoid overfitting, a condition where the neural network has a much lower performance in the test database than the training and/or validation database. Equations 6 and 7 demonstrate the mean squared error (MSE) and the mean absolute error (MAE) respectively.

$$MSE = \frac{1}{N} \sum \left(y_{true} - y_{pred}\right)^2 \tag{6}$$

$$MAE = \frac{1}{N} \sum \left|y_{true} - y_{pred}\right| \tag{7}$$

$N$ is the amount of data of the target variable, $y_{true}$ is the value of the target variable and $y_{pred}$ is the predicted value of the target variable.

## 2.3 Artificial neural network Statistical analysis

Sensitivity analysis is used to establish the contribution of each predictor variable in an artificial neural network [43]. There are techniques to perform sensitivity analysis in artificial neural networks, such as Garson's algorithm, partial derivatives, input perturbation, forward stepwise addition, backward stepwise elimination [44] and the quotient W [45]. The chosen sensitivity analysis was the connection weights algorithm [43] as it demonstrates the highest Gower's coefficient of similarity in 500 Monte Carlo simulations [44] and denotes the similarity between the true ranked importance and estimated ranked importance of the variables in the artificial neural network [46].

In the statistical evaluation stage of the developed artificial neural network, the metrics of mean absolute error (MAE), standard deviation of error (Std) and the coefficient of determination ($R^2$) were used. The standard deviation of error and the coefficient of determination can be seen in Equation 8 and 9, respectively. To compare the results of the artificial neural network with models from the literature, the equations Shaw, Watt-Fereday, Bomkamp, Riboud, Duchesne and ANNliq, detailed in Duchesne et al. [17], and the Viscosity module of FactSage[TM] 7.2 [47] were used. The skewness and kurtosis of the deviations of the best models were analyzed to evaluate and indicate the variability of the predictor models.

$$\text{Std} = \sqrt{\frac{1}{N-1} \sum (\alpha - \mu_\alpha)^2} \qquad (8)$$

$$R^2 = 1 - \frac{\sum (y_{true} - y_{pred})^2}{\sum (y_{true} - \mu_{true})^2} \qquad (9)$$

$\alpha = |y_{true} - y_{pred}|$, $\mu_\alpha$ is $\alpha$ mean e $\mu_{true}$ is the target variable mean.

Table 3 — Hyperparameters used in training of artificial neural network.

| Hiperparameter | Experiments | Selected |
|---|---|---|
| Input layer | 2, 8, 20 | 20 |
| Hidden layers | 1, 2, 3, 4, 5, 6, 7, 8, 9, 10, 11, 12, 13, 14, 15, 16, 17, 18, 19, 20, 21, 22 | 22 |
| Output layer | 1 | - |
| Depth | 1, 2, 3 | 3 |
| Activation function | Sigmoid, Tanh*, ReLU | ReLU |
| Weight initialization | He uniform [41] | - |
| Bias initialization | Zeros, Ones | Zeros |
| Loss | Mean Squared Error | - |
| Metrics | Mean Absolute Error | - |
| Optimizer | Adam [42] | - |
| Epochs | 10000, 50000, 100000 | 100000 |
| Batch size | 2, 4, 8, 16, 32, 64 | 64 |
| Early Stopping | True | - |

* Hyperbolic tangent.

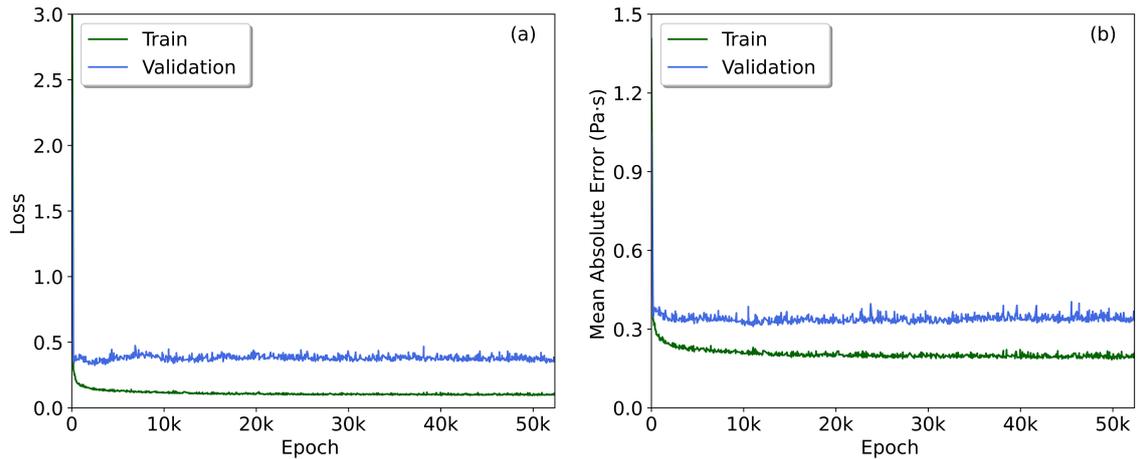

Figure 3 — *Loss*x*Epoch* (a) and *MAE*x*Epoch* (b) in Viskositas.

## 3 Results and discussion

### 3.1 Artificial neural network training and validation

The artificial neural network that presented the best result was lower loss and lower metrics obtained 3 hidden layers. The neural network with the best hyperparameters was named *Viskositas* with 20 input data (chemical composition and temperature), 3 hidden layers with 22 neurons each and the output layer with 1 neuron, the $\log_{10}$ of viscosity. Figure 3 (a) demonstrates the loss variation by epoch during the training and validation phase of Viskositas. In Figure 3 (a) it is possible to analyze that the maximum number of selected epochs (100000) was not obtained because during the training and validation phase the EarlyStopping technique was used, avoiding overfitting and helping to generalize the model. Figure 3 (b) demonstrates the relationship between the metrics given by the mean absolute error (MAE) and the epochs during the Viskositas training and validation.

In Figure 3 (a) and (b) can be seen that the validation step has higher loss and higher MAE in relation to the training, which is a usual behavior in neural networks [16, 18-19]. Graphs are types of Euclidean and non-Euclidean data structure capable of modeling objects and their relationships that can be used as denotation of a large number of systems in different areas [30]. The Viskositas model graph can be seen in Figure 4.

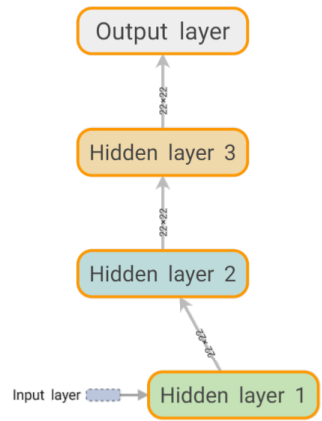

Figura 4 — Viskositas model graph.

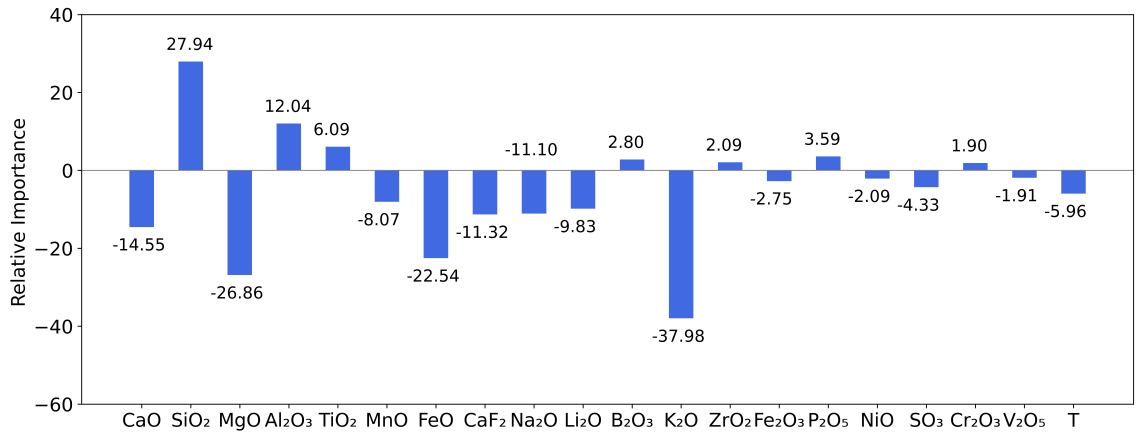

Figure 5 — Relative importance of each Viskosite input data (T = Temperature).

## 3.2 Sensitivity analysis

Applying the connection weights algorithm, the relative importance was obtained in relation to each input data of Viskositas. Figure 5 demonstrates the relative importance of chemical composition and temperature in relation to predicted viscosity. When a predictor variable has negative relative importance it means that it is inversely proportional to the target variable and when a predictor variable has positive relative importance means that this variable is directly proportional to the target variable. A higher absolute numerical value of the relative importance of a predictor variable indicates greater sensitivity to the target variable and a lower absolute numerical value of the predictor variable indicates less sensitivity to the target variable [48]. Considering the chemical system of the composition of the Viskositas input data, analyzes were carried out on the behavior via chemical formula and in relation to temperature.

### 3.2.1 MO oxides

MO oxides (e.g. CaO, MgO, MnO, FeO, NiO) are considered modifier species because they have the ability to weaken chains between network-forming species. A reaction mechanism between MO species presence with silicates can be seen in Equation 10 [7].

$$\equiv Si\text{-}O\text{-}Si\equiv \;+\; MO \rightarrow \;\equiv Si\text{-}O\text{-}M\text{-}O\text{-}Si\equiv \;+\; O^{2-} \qquad (10)$$

With the dissociation in the chemical species MO, there is a reaction between the $M^{2+}$ ions that weaken the chain between Si-O-Si. MgO decreases the viscosity of chemical systems $SiO_2$-CaO-$Al_2O_3$-MnO-MgO-$K_2O$ [49], FeO decreases the viscosity of chemical systems CaO-$SiO_2$-$Al_2O_3$-MgO-$P_2O_5$ [50], CaO and MgO decay the viscosity of chemical systems CaO-MgO-$SiO_2$-$Al_2O_3$ [51] and CaO-$SiO_2$-$Al_2O_3$-$B_2O_3$-$Na_2O$-$TiO_2$-MgO-$Li_2O$-MnO-$ZrO_2$ [52], MnO in binary systems $SiO_2$-MnO [53] and the addition of NiO decreases the viscosity of chemical systems $SiO_2$-$Al_2O_3$-$Fe_2O_3$-CaO-MgO-$TiO_2$-$SO_3$-$K_2O$-$Na_2O$-$P_2O_5$ [54] in fully liquid physical systems. The addition of MO chemical species can cause an increase in viscosity in chemical systems when linked to the precipitation of solid phases, thus increasing their viscosity by the non-Newtonian behavior indicated by the Roscoe-Einstein equation [55] (Equation 11).

$$\eta_{eff} = \eta\,(1 - ac)^b \qquad (11)$$

$\eta_{eff}$ is the effective viscosity in liquid-solid system, $\eta$ the liquid viscosity, $c$ the volume concentration of solids, $a$ and $b$ system-dependent constants [55-57].

### 3.2.2 M$_2$O oxides

$Na_2O$, $Li_2O$ and $K_2O$ are oxides composed of alkaline metals capable of complete dissociation and weakening the silicate chain and other network formers, thus defined as network modifiers. Equation 12 [7] demonstrates the action of $M_2O$ oxides on a silicate network.

$$\equiv Si\text{-}O\text{-}Si\equiv \;+\; M_2O \rightarrow \;\equiv Si\text{-}O\text{-}M \;+\; M\text{-}O\text{-}Si\equiv \qquad (12)$$

Equation 12 can be described as the sum of 3 half-reactions, indicated by Equations 13-15.

$$M_2O \rightarrow 2M^+ + O^{2-} \qquad (13)$$

$$\equiv Si\text{-}O\text{-}Si\equiv \;+\; O^{2-} \rightarrow 2\equiv Si\text{-}O^- \qquad (14)$$

$$2\equiv Si\text{-}O^- + 2M^+ \rightarrow \;\equiv Si\text{-}O\text{-}M \;+\; M\text{-}O\text{-}Si\equiv \qquad (15)$$

Equation 14 denotes an implicit reaction concerning the reaction between free oxygen (FO; $O^{2-}$), bridging oxygen (BO; $O^0$) and non-bridging oxygen (NBO, $O^-$) (Equation 16) [7-49].

$$O^{2-} + O^0 \rightarrow 2O^- \qquad (16)$$

The action of $O^{2-}$ also results from the depolymerization of chains of chemical systems. $Na_2O$ decreases the viscosity of chemical systems CaO-$SiO_2$-$Al_2O_3$-$B_2O_3$-$Na_2O$-$TiO_2$-MgO-$Li_2O$-MnO-$ZrO_2$ [52], $Li_2O$ decreases the viscosity of chemical systems $SiO_2$-CaO-$Al_2O_3$-MgO-$F^-$-$Na_2O$-MnO-$Li_2O$-$B_2O_3$ [58] and the action $K_2O$+$Na_2O$ decreases the viscosity of fluxes [59].

### 3.2.3 M$_2$O$_3$ oxides

The chemical species that behaves as a network modifier and network former depending on the chemical system where it is present is called amphoteric (e.g. $Al_2O_3$, $B_2O_3$, $Fe_2O_3$, $Cr_2O_3$). When $M_2O_3$ acts as a network former, there is an electrical charge compensation by the action of an ion with valence number +1, as indicated by Equation 17 [7] with the participation of $Al_2O_3$.

$$2\equiv Si\text{-}O\text{-}Si\equiv + Al_2O_3 + M^+ \rightarrow 2\equiv Si\text{-}O\text{-}Al^{M+}\equiv \qquad (17)$$

The $M_2O_3$ oxides weaken the structure of chemical systems by presenting high coordination [7] and by reacting with $O^{2-}$ [49] for example. $B_2O_3$ decreases the viscosity of chemical systems $CaO\text{-}SiO_2\text{-}Al_2O_3\text{-}B_2O_3\text{-}Na_2O\text{-}TiO_2\text{-}MgO\text{-}Li_2O\text{-}MnO\text{-}ZrO_2$ [52], $Fe_2O_3$, $Cr_2O_3$ and $Al_2O_3$ have amphoteric behavior in chemical systems $Fe_2O_3\text{-}SiO_2\text{-}FeO\text{-}MgO\text{-}Al_2O_3\text{-}CaO$ [60], $CaO\text{-}SiO_2\text{-}MgO\text{-}Al_2O_3\text{-}TiO_2\text{-}Cr_2O_3$ [61] and $Al_2O_3\text{-}Na_2O\text{-}K_2O$ [62] respectively. $Al_2O_3$ increases the viscosity of chemical systems $Al_2O_3\text{-}MgO$ [62] and $SiO_2\text{-}CaO\text{-}Al_2O_3\text{-}MgO\text{-}F^-\text{-}Na_2O\text{-}MnO\text{-}Li_2O\text{-}B_2O_3$ [58]

### 3.2.4 $M_2O_5$ oxides

Also considered to form networks, the $M_2O_5$ species (e.g. $P_2O_5$, $V_2O_5$) usually increase the viscosity of chemical systems, and why is not clear [63]. The addition of $P_2O_5$ increases the viscosity of $CaO\text{-}SiO_2\text{-}Al_2O_3\text{-}MgO\text{-}TiO_2$ chemical systems [63]. The action of $V_2O_5$ may result from the decrease in the viscosity of chemical systems $SiO_2\text{-}Al_2O_3\text{-}Fe_2O_3\text{-}CaO\text{-}MgO\text{-}TiO_2\text{-}SO_3\text{-}K_2O\text{-}Na_2O\text{-}P_2O_5$ [54] by interaction between V and Al atoms that destroy the structure, reducing its viscosity [54].

### 3.2.5 $MO_2$ oxides

Network formers are chemical species capable of forming coordinated tetrahedrons, increasing the degree of polymerization of the chemical system (e.g. $SiO_2$, $TiO_2$, $ZrO_2$). O $SiO_2$ is a fundamental species for understanding the structures of chemical systems. The NBO/T of silicates can be divided between 0, 1, 2, 3 and 4 presenting the structural parameters $Si_2O_8^{8-}$, $Si_2O_7^{6-}$, $Si_2O_6^{4-}$, $Si_2O_5^{2-}$ and $Si_2O_4$ respectively [64]. The representation of the structural parameters can be done by the notation $Q^n$ where $n$ is the number of bridging oxygen [65] corresponding to the pair $Q^0\text{-}SiO_4^{4-}$, $Q^1\text{-}Si_2O_7^{6-}$, $Q^2\text{-}SiO_3^{2-}$ e $Q^3\text{-}Si_2O_5^{2-}$ [66]. The interaction between structural parameters $Q^n$ can modify the structure of a chemical system, as demonstrated by Equation 18.

$$2Q^n \rightarrow Q^{n-1} + Q^{n+1} \qquad (18)$$

$SiO_2$ increases the viscosity of $SiO_2\text{-}FeO$ [53], $SiO_2\text{-}Al_2O_3$ [62] and $CaO\text{-}Fe_2O_3$ [68] chemical systems when added, $TiO_2$ increases the viscosity of $CaO\text{-}SiO_2$ and $CaO\text{-}Al_2O_3$-based mold flux [69]. $ZrO_2$ does not have much influence on viscosity [59].

### 3.2.6 Fluorite

Fluorite ($CaF_2$) is also considered a chemical species that decreases the viscosity of chemical systems. Two mechanisms of fluorite in silicates are described in Equations 19-20 [70-72]. There is a break in the structure by the action of fluorite, thus decreasing the viscosity of the chemical system. Fluorite decreases the viscosity of the chemical systems $SiO_2\text{-}Al_2O_3\text{-}MgO\text{-}MnO_2\text{-}Fe_2O_3\text{-}Na_2O\text{-}CaO\text{-}K_2O\text{-}CaF_2$, $CaO\text{-}SiO_2\text{-}Al_2O_3\text{-}CaF_2$ [70], $CaO\text{-}SiO_2\text{-}MgO\text{-}CaF_2$ [71], $CaO\text{-}SiO_2\text{-}Na_2O\text{-}Al_2O_3\text{-}MgO\text{-}CaF_2\text{-}FeO\text{-}Li_2O\text{-}B_2O_3$ [72], $CaO\text{-}SiO_2$, $CaO\text{-}Al_2O_3$-based mold flux [69] and in fluxes [59].

$$Si_3O_9^{6-} + 2F^- \rightarrow Si_2O_6F^{5-} + SiO_3F^{3-} \qquad (19)$$

$$Si_2O_6F^{5-} + SiO_3F^{3-} + 2F^- \rightarrow 2SiO_3F^{3-} + SiO_2F_2^{2-} + O^{2-} \qquad (20)$$

### 3.2.7 Temperature

The Vogel-Fulcher-Tammann (VFT) equation correlates the relationship between viscosity and temperature, as described in Equation 21 [24].

$$\log \eta = a + \frac{b}{T-c} \tag{21}$$

$\eta$ is the viscosity, $a$, $b$ e $c$ constants and $T$ the temperature. The Arrhenius equation is a simplified equation of the VFT equation, where the constant c=0. The temperature is inversely proportional to the viscosity by decreasing the degree of agitation between the chemical species in its decrease, decreasing its fluidity and thus increasing the viscosity. The opposite is also true. When the temperature increases, the viscosity of chemical systems $SiO_2$-$CaO$-$Al_2O_3$-$MnO$-$MgO$-$K_2O$ [49], $CaO$-$SiO_2$-$Al_2O_3$-$MgO$-$P_2O_5$ [50], $CaO$-$MgO$-$SiO_2$-$Al_2O_3$ [51], $Fe_2O_3$-$SiO_2$-$FeO$-$MgO$-$Al_2O_3$-$CaO$ [60], $CaO$-$SiO_2$-$MgO$-$Al_2O_3$-$TiO_2$-$Cr_2O_3$ [61], $Al_2O_3$-$Na_2O$-$K_2O$ [62], $CaO$-$SiO_2$-$Al_2O_3$, $CaO$-$Fe_2O_3$ [68], $SiO_2$-$Al_2O_3$-$MgO$-$MnO_2$-$Fe_2O_3$-$Na_2O$-$CaO$-$K_2O$-$CaF_2$, $CaO$-$SiO_2$-$Al_2O_3$-$CaF_2$ [70], $CaO$-$SiO_2$-$MgO$-$CaF_2$ [71] e $CaO$-$SiO_2$-$Na_2O$-$Al_2O_3$-$MgO$-$CaF_2$-$FeO$-$Li_2O$-$B_2O_3$ [72] decreases.

### 3.3 Statistical Analysis

The results of MAE, Std and $R^2$ of models Shaw, Watt-Fereday, Bomkamp, Riboud, Duchesne, ANNLiq, the Viscosity module of FactSage[TM] 7.2 and the Viskositas model can be seen in Appendix.

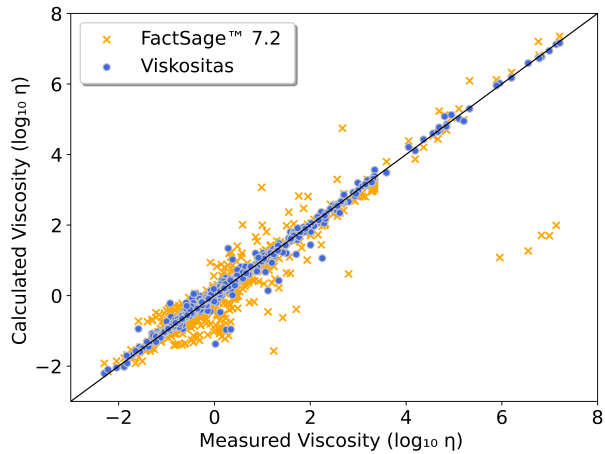

Figure 6. Relationship of deviations of the Viskositas model and the Viscosity module of FactSage[TM] 7.2 in relation to the test database.

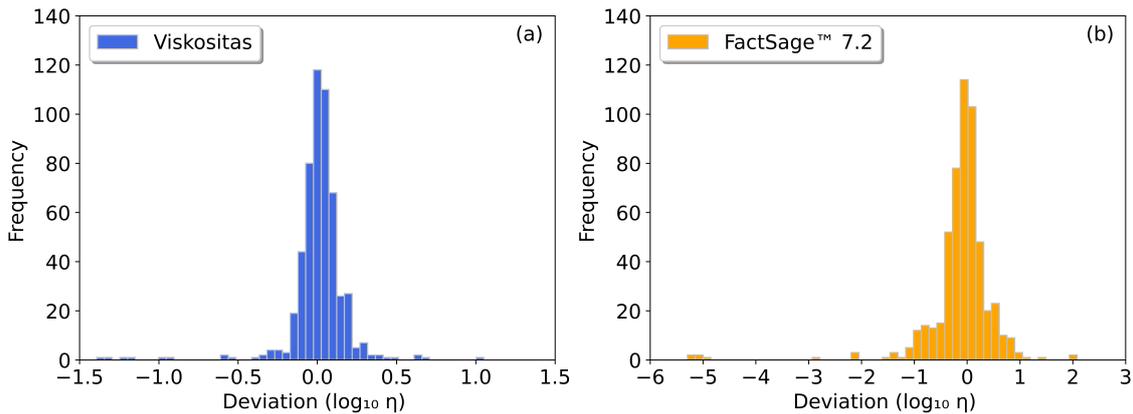

Figure 7. Histograms representing the deviations of the Viskositas model (a) and the Viscosity module of FactSage[TM] 7.2 (b) in relation to the test database.

The best results for MAE, Std and $R^2$ were presented by the Viskositas model and by the Viscosity module of FactSage[TM] 7.2. The relationship between the test data and the data predicted by the Viskositas model and by the Viscosity module FactSage[TM] 7.2 can be seen in Figure 6. The Viskositas model has lower MAE and Std because it has lower error and lower dispersion and higher $R^2$ in relation to the database of test data.

The histogram of the deviations of the Viskositas model and the Viscosity module of FactSage™ 7.2 can be seen in Figure 7. The skewness and kurtosis of the deviations of the Viskositas model and the Viscosity module of FactSage™ 7.2 were also evaluated, with the maximum error with respect to positive and negative asymmetry (Table 4). The Viskositas model showed lower skewness, kurtosis, maximum errors in relation to positive and negative asymmetry, demonstrating lower generation of outliers [73] in predicting the viscosity of multicomponent chemical systems.

Table 4. Values of skewness, kurtosis, maximum error in relation to positive and negative asymmetry of the Viskositas model (a) and of the Viscosity module of FactSage™ 7.2 (b) in relation to the test database.

| Analysis | Skewness | Kurtosis | Max. Negative error ($\log_{10} \eta$) | Max. Positive error ($\log_{10} \eta$) |
|---|---|---|---|---|
| Viskositas | -2.6059 | 22.1878 | -1.3920 | 1.0507 |
| Viscosity (FactSage™ 7.2) | -4.0932 | 28.9009 | -5.2931 | 2.0832 |

# 4 Conclusion

This article analyzed the importance of viscosity in different chemical systems, applicable in different branches of industry, considered a variable that is difficult to measure. Using a literature database, a nonlinear model was built using artificial neural networks by variation of hyperparameters. The model built was named *Viskositas* and showed lower mean absolute error (MAE), standard deviation (Std) and higher coefficient of determination ($R^2$) in relation to the test database when compared to 6 different models from the literature and 1 commercial model, also demonstrating lower generation of outliers.

# APPENDIX

Table 5. Results of MAE, Std and $R^2$ of models Shaw, Watt-Fereday, Bomkamp, Riboud, Duchesne, ANNLiq, the Viscosity module of FactSage$^{TM}$ 7.2 and the Viskositas model in test data.

| Model | Shaw | Watt-Federay | Bomkamp | Riboud | Duchesne | ANNLiq | Viscosity (FactSage$^{TM}$ 7.2) | *Viskositas* |
|---|---|---|---|---|---|---|---|---|
| Mean Absolute Error (MAE) ($\log_{10} \eta$) | 1.4217 | 1.5558 | 3.9684 | 2.8301 | 2.8027 | 11.6147 | 0.8112 | 0.2309 |
| Standard Deviation (Std) ($\log_{10} \eta$) | 1.6857 | 1.8218 | 2.2953 | 1.9337 | 2.3038 | 4.8249 | 1.3348 | 0.5446 |
| Coefficient of Determination ($R^2$) | 0.6452 | 0.6068 | 0.5395 | 0.5297 | 0.1611 | 0.1419 | 0.8212 | 0.9864 |